\documentstyle[12pt]{article}
\title{\bf Dynamically tuning away the cosmological constant \\
in effective scalar tensor theories}
\vspace{2cm}
\author {Nimmi Rooprai\thanks{E--mail : nimmi@ducos.ernet.in},
Daksh Lohiya\thanks{E--mail : dlohiya@ducos.ernet.in}\\
       {\em Department of Physics and Astrophysics}\\
       {\em  University of Delhi, Delhi-110 007, India}
       }
\begin{document}
\renewcommand{\today}{}
\maketitle
\begin{center}
\large{\bf Abstract}
\end{center}

It is known that the cosmological constant can be dynamically tuned to
an arbitrary small value in classes of scalar tensor theories.  The
trouble with such schemes is that effective gravity itself vanishes.
We explore the possibility of avoiding this ``no-go'' with a spatially
varying effective gravity.  We demonstrate this in principle with the
non-minimally coupled scalar field having an additional coupling to a
fermionic field. The expectation value of the scalar field gets
anchored to a non-trivial value inside compact domains.  But for the
non-minimal coupling to the scalar curvature, these configurations are
analogous to the non-topological solutions suggested by Lee and
Wick. With non-minimal coupling, this leads to a peculiar spatial
variation of effective gravity. As before, one can dynamically have
the long distance (global) gravitational constant $G$ and ${\Lambda}$,
the cosmological constant, tending to zero. However, inside compact
domains, $G$ can be held to a universal (non-vanishing) value.  Long
distance gravitational effects turn out to be indistinguishable from
those expected of general theory of relativity (GTR). There are two
ways in which the ensuing theory may lead to a viable effective
gravity theory: (a) the compact domains could be of microscopic
(sub-nuclear) size, or (b) the domains could be large enough to
accommodate structures as large as a typical galaxy.  Aspects of
effective gravity and cosmology that follow are described. A toy
Freidman - Robertson - Walker (FRW) model free from several standard
model pathologies and characteristic features emerges.

\pagebreak

\begin{section}*{1. Introduction:}

It is now widely accepted that a theory of gravitation described by
the Einstein - Hilbert action:
\begin{equation}
S = \int d^4x \sqrt{-g}[{1\over {8\pi G}}R + \Lambda]
\end{equation}
fares wonderfully well to classical precision tests. The success is
impressive enough to rule out most alternative theories at the
classical level \cite{will}.  However, the theory described by eqn(1)
has severe theoretical inconsistencies.  The appearance of a
dimensional gravitational coupling constant $G$, and the smallness of
the cosmological constant $\Lambda$, are two pathologies that impede
any attempt to treat this action as a fundamental quantum theory.  The
dimensional coupling is primarily responsible for the non -
renormalizability of the theory \cite{wein1} while the smallness of
$\Lambda$ can not be justified in any reasonable theoretical model
\cite{wein2} Besides these theoretical problems, a cosmological model
based on the above action has to contend with dynamical
inconsistencies: reasonable initial conditions can not be dynamically
realized on account of the horizon and fine - tuning (flatness)
problems. A resolution of these dynamical inconsistencies using
inflation, without addressing the cosmological constant problem is, at
best, incomplete. Inflation is founded upon the possibility that the
early universe witnessed several phase transitions - each producing a
large change in the effective value of the cosmological constant. Be
it so, the last transition is required to exit to a state of a tiny
cosmological constant. All versions of inflation require fine tuning
of model parameters to get inflation going and / or to get inflation
exit gracefully to the standard early hot universe \cite{linde}.
The least troublesome version of inflation, the chaotic
model, assigns to $\Lambda$ a vanishing value by hand. Anthropic
arguments are invoked to justify a small value for $\Lambda$
\cite{wein3,vilen,steinh} and have invited expressions of despondency
and a need for a better ansatz. We share Witten's \cite{witten}
reservations on the use of anthropic arguments in general.

Over the last two decades there is an emerging consensus that the
Einstein - Hilbert action ought to be regarded, at most, as a
correct low energy limit of a decent quantum theory. Superstring
theory has emerged as a promising candidate in this respect. It is
hoped that an effective gravity theory would emerge from the dynamics
of an appropriate string theory.  A lack of understanding of the
vanishing (or the tininess) of the cosmological constant is the key
obstacle to making realistic string-inspired particle physics
models. A convincing mechanism of supersymmetry breaking has not
emerged as yet. All known approaches generate a large cosmological
constant and must therefore be regarded as incorrect \cite{witten}.

On the other hand, general considerations regarding the structure of
string theory suggest that general relativity may acquire
modifications, even at energies considerably lower than the Planck
scale.  Dilaton gravity theories in ten dimensions have recently
played a prominent role in a search for a consistent effective low
energy limit emerging from string inspired quantum theory of gravity
\cite{pol}.  This has resulted in a resurrection of interest in scalar
- tensor and Brans-Dicke models.  Such models have a characteristic
coupling of the ricci scalar curvature with a function of a scalar
field - defining a non - minimal coupling (NMC).  A search for a
suitable compactification scheme that would yield the Einstein -
Hilbert action in four dimensions is on within the framework of such
models. To achieve this, the foremost requirement is to have a
provision for an appropriate fixing of NMC to give a universal
gravitational constant.

The NMC could, for example, be determined by the minimum of an
effective potential of the scalar field. Such a potential may arise as
a result of the compactification scheme itself or as a result of
quantum corrections \cite{afflek}. Determining the NMC by the minimum
of an effective potential leads to a realization of a theory
indistinguishable from the Einstein - Hilbert theory at low energies
\cite{Zee}. However, such a scheme falls short of addressing itself to
the cosmological constant problem as again, there is nothing to ensure
the tininess of the minimum value of the effective potential.

There have  been attempts to {\it dynamically} tune the effective
cosmological constant to a vanishingly small value in classes of NMC
theories \cite{dolgov,ford}.  All such attempts are frustrated by a
``no-go theorem'' \cite{wein2}.  In these schemes, the scalar field
develops an instability, the non-minimal coupling diverges and the
scalar field stress tensor evolves to neutralize the effective value
of the cosmological constant to zero. The trouble with such a scheme
is that with the diverging NMC, the effective gravitational constant
itself vanishes.  This no-go situation has not been successfully
evaded.  One has not found a natural way to stabilize the NMC to a
bounded value that could be identified with $\approx (8\pi G)^{-1}$
and, at the same time, tune the cosmological constant to a small
value.

In a sweeping review, Weinberg \cite{wein2} has made out a case for
the use of a judicious anthropic argument as an {\it explanation} for
the tininess of the cosmological constant. One considers formalisms
that allow an effective cosmological constant to appear dynamically in
the action and appeals to some unique feature that would pick out a
small value for this constant in the functional integral. The most
explored principal is the imposition of a conditional probability on
the functional integral that would allow for a large enough universe
to exist and also allow for gravitational clustering at a sufficiently
small redshift in the universe. This imposes severe constraints on the
cosmological constant.  Indeed the most recent attempts
\cite{wein3,vilen,steinh} to justify a small $\Lambda$ uses an
anthropic argument to have a vanishing minimum to an effective
``tracker'' potential for a scalar field.  The peculiar profile of
such potentials, and a slow asymptotic rolling of the scalar field
near the minimum, can in principle allow for an effective $\Lambda$
dominated cosmology at the current epoch.

A feature that would pick out a small cosmological constant using an
objective physical principle rather than having to take a recourse to
an anthropic principle, is worth exploring. In this article we
describe solutions in a class of non-trivial NMC theory having a {\it
{minimum}} energy at a vanishing cosmological constant. The solutions
arose in our search for an ansatz that would allow for a NMC with a
non-trivial spatial dependence. As a result of a scalar field coupling
to a fermion field, the expectation value of the scalar field can
acquire a non-trivial value inside compact domains.  But for the
non-minimal coupling (of the scalar field) to the scalar curvature,
these configurations are similar in character to the non-topological
solutions suggested by Lee and Wick \cite{wick}.  With NMC, one gets a
peculiar spatial variation of effective gravity. We shall require $G$
to be held to a non-trivial value inside compact domains.  This is
sufficient to have long distance gravitational effects
indistinguishable from those expected from general theory of
relativity (GTR). We point out two ways this could be possible: (a)
the domains could be large enough to accommodate structures as large
as a typical galaxy or (b) the compact domains could be of microscopic
(sub-nuclear) size. This is described in the next section where we
also describe some essential aspects of a generalized effective scalar
tensor theory. The action is chosen to include scalar functions
multiplying the ricci scalar as well as a fermionic part.  Stability
of large fermion number non-trivial solutions requires their interior
to be near the zero of the effective potential.

In the last section we discuss features of a toy cosmological model
that such a theory can support. The struture has definitive,
falsifiable predictions. In standard cosmology, one has pinned down
hopes on special initial conditions, motivated by anthropic
considerations to account for the observed cosmological constant. Such
tautologies, when applied to the universe as a whole, have little
predictive power.  In the structure described in this article, any
fine tuning of coupling parameters of the fermion lagrangian that may
be required to have stable gravity domains of required size, has
testable consequences.
\end{section} 

\begin{section}*{2. A generalized scalar tensor theory:}
\begin{subsection}*{2(a). The action and field equations:}
Consider a classical theory of a scalar field $\phi$ coupled to the
scalar curvature $R$, through an arbitrary function $U(\phi)$, in the
action:
\begin{equation}
\label{eq:lag}
S = \int d^4x\sqrt{-g}[U(\phi)R 
+ {1\over 2}\partial^\mu\phi\partial_\mu\phi
- V(\phi) + L_m] 
\end{equation}
We shall treat $S$ as a classical, effective, phenomenological action
that could be good enough to do tree level (``on-shell'') physics. 
If need arises, we shall consider non polynomial forms for the effective
potential and the NMC: $V(\phi)$ and $U(\phi)$.  Such profiles have
been used as ``tracker-field'' potentials \cite{wein3, steinh} and can
be justified in particle physics models with dynamic symmetry breaking
or non-perturbative effects.  In particular they can arise from
non-perturbative effects that lift flat directions in supersymmetric
gauge theories as well as in modulii fields in string theory
\cite{afflek, witt} (see also \cite{wett}).  Strictly speaking, we
would consider it premature to justify the forms for these functions
at this formative stage on the basis of fundamental physics. For our
purpose, we simply regard eqn(\ref{eq:lag}) as the action of a 
parametrized Brans-Dicke fluid.  Any arbitrary (cosmological) constant
in the theory is included in $V(\phi)$.  $L_m \equiv L_w + L_\psi$ is
the contribution from the rest of the matter fields. It includes a
contribution from a Dirac fermion field:
\begin{equation}
\label{eq:dirac}
L_\psi 
\equiv \bar{U}(\phi)^{-1}
[{1\over 2}[\bar{\psi}\overleftarrow{D}_\mu\gamma^\mu\psi
- \bar{\psi}\gamma^\mu\overrightarrow{D}_\mu\psi]
- m(\phi)\bar{\psi}\psi] 
\end{equation}
Here $D_\mu$ is the spin covariant derivative \cite{pa}.
This Fermion action has an overall scaling function
$\bar U(\phi)^{-1}$. Such an arbitrary scaling
can be absorbed in a redefinition of the Dirac field:
$\psi\longrightarrow \psi' \equiv \psi/\bar U^{1\over 2}$ provided
$\bar U$ were bounded. Indeed, by such a scaling (for a ${\it{bounded}}~~
\bar U$), the $\phi$ dependence in $L_\psi$ can be constrained to the
mass munction $m(\phi)$. The anti-symmetrized derivative appearing
in the fermion lagrangian ensures the cancellation of the derivative
of a real, non-vanishing, bounded scaling function. 
\footnote{This result is
not specific to the chosen form of the Dirac Lagrangian. 
Covariant spinor fields  of arbitrary weights can be
defined as geometrical objects over (direct products of) 
two dimensional complex
spaces. An arbitrary, non vanishing, real scaling of such a spinor can
be absorbed into a real part of the trace of a spinor connection and
does not contribute to the covariant derivative of the spinor and therefore
has no interaction with other fields \cite{Peres}.} 
If $\bar U(\phi)^{-1}$
vanishes at any point, the Dirac action identically vanishes. There is
no dynamics for the Dirac field at such a point.  The invariance of
the Dirac Lagrangian under arbitrary phase change: $\delta\psi =
i\epsilon\psi$ implies the vanishing divergence of the current:
$$
J^\mu = \bar U(\phi)^{-1}\bar\psi\gamma^\mu\psi \eqno{(4)}
$$
and the conservation of the charge:
$$
Q \equiv \int_\Sigma \bar U(\phi)^{-1}\bar\psi\gamma^o\psi \eqno{(5)}
$$
The Dirac particles would be confined to regions where $\bar U$ is
bounded i.e.  $\bar U(\phi)^{-1}$ is non - vanishing.  The only effect
of a scaling function $\bar U(\phi)^{-1}$ that vanishes outside a
compact domain is to confine the fermions within the domain. Inside
the domain the scaling can be identically absorbed in a rescaling of
the fermion field. In what follows we shall essentially do this.  This
leads to conditions similar to those that arise in non-topological
soliton solutions in the Lee-Wick model \cite{wick}. {\it One could
just as well have followed the Lee-Wick model in which confinement
occurs for fermions that are not on-shell outside a domain on account
of a large coupling with the Higgs field.}

Our foremost task would be to demonstrate the vanishing of the
covariant divergence of the stress tensor of the rest of the matter
fields described by $L_w$ in eqn(\ref{eq:lag}).  Requiring the action to be
stationary under variations of the metric tensor and the fields $\phi
~\&~ \psi$ gives the equations of motion:
$$
U(\phi)[R^{\mu\nu} - {1\over 2}g^{\mu\nu}R] = -{1\over 2}[T_w^{\mu\nu} 
+ T_\phi^{\mu\nu} + \Theta^{\mu\nu}   
+ 2U(\phi)^{;\mu;\nu} 
- 2g^{\mu\nu}U(\phi)]^{;\lambda}_{;\lambda}] \eqno{(6)}
$$
$$
g^{\mu\nu}\phi_{;\mu;\nu} + {\partial V\over {\partial\phi}} 
- R{\partial U \over {\partial \phi}} + 
{\partial m\over {\partial \phi}}\bar\psi'\psi'  
= 0 \eqno{(7)}
$$
$$
\gamma^\mu D_\mu\psi' + m(\phi)\psi' = 0\eqno{(8)}
$$
$$
D_\mu\bar\psi'\gamma^\mu - m(\phi)\bar\psi' = 0\eqno{(9)}
$$
Here $T_w^{\mu\nu}$, $\Theta^\mu_\nu$
are the energy momentum tensors constructed from
$L_w$ and $L_\psi$ respectively, and
$$
T_\phi^{\mu\nu} \equiv  \partial^\mu\phi\partial^\nu\phi 
- g^{\mu\nu}[{1\over 2}\partial^\lambda\phi\partial_\lambda\phi -V(\phi)]
\eqno{(10)}
$$
$L_w$ is taken to be independent of $\phi$ and $\psi$.  
The two Dirac
equations (8) and (9) ensure $L_\psi$ to be null. 
The scalar field equations are
therefore independent of $\bar U(\phi)$.  The Fermion stress tensor is
simply:
$$
\Theta^\mu_\nu \equiv 
-{1\over 2}[\bar\psi'\overleftarrow{D}_\nu\gamma^\mu\psi'
- \bar\psi'\gamma^\mu\overrightarrow{D}_\nu\psi']\eqno{(11)}
$$
The covariant divergence of (6) is easily seen to reduce to:
$$
\Theta^\mu_{\nu;\mu} = 
{\partial m\over{\partial\phi}}\partial_\nu\phi\bar\psi'\psi'
\eqno{(12)}
$$
Thus there is a violation of equivalence principle as far as the Dirac
field is concerned. However, in a region where the scalar field
gradient vanishes or for a $\phi$ independent fermion mass, the
covariant divergence of the fermion field stress tensor vanishes. To
see how the equivalence principle strictly holds for the rest of the
matter fields, i.e.: $T^{\mu\nu}_{w;\nu} = 0$, consider the covariant
divergence of eqn(6). From the contracted Bianchi identity satisfied by
the Einstein tensor, we obtain
$$
U(\phi)_{,\nu}[R^{\mu\nu} - {1\over 2}g^{\mu\nu}R] = 
-{1\over 2}[T^{\mu\nu}_{w;\nu} + t^{\mu\nu}_{;\nu}
+ \Theta^{\mu\nu}_{;\nu}] \eqno{(13)}
$$
with 
$$t^{\mu\nu} \equiv T_\phi^{\mu\nu} +  2U(\phi)^{;\mu;\nu} -
2g^{\mu\nu}U(\phi)^{;\lambda}_{;\lambda} \eqno{(14)}$$
Using the identity: 
$$U(\phi)^{;\rho}R_{\rho\alpha} = U(\phi)^{;\lambda}_{;\lambda ;\alpha}
- U(\phi)^{;\lambda}_{;\alpha;\lambda}$$ 
and  eqn(12), eqn(13) reduces to
$$
-{1\over 2}U(\phi)^{,\mu}R = 
-{1\over 2}[T^{\mu\nu}_{w;\nu} + T^{\mu\nu}_{\phi;\nu}
+ \partial^\mu m\bar\psi'\psi'] \eqno{(15)} 
$$
Finally, using the equation of motion for the scalar field (7),
all the $\phi$ dependent terms cancel the left hand side -
giving the vanishing of the covariant divergence of the (w-) matter
stress energy tensor. This in turn ensures that equivalence principle holds
- in turn assuring the geodesic law.
\end{subsection}
\begin{subsection}*{2(b). The conserved energy:}

Next, we write down the expression for a conserved pseudo energy
momentum tensor \cite{dlsb}. Defining
$$
{\rm A} \equiv
\sqrt{-g}g^{\sigma\rho}[\Gamma^\alpha_{\sigma\rho}\Gamma^\beta_{\alpha\beta}
- \Gamma^\alpha_{\beta\rho}\Gamma^\beta_{\alpha\sigma}]\eqno{(16)}
$$
$$
{\rm B} \equiv [U{\rm A} 
- \sqrt{-g}g^{\sigma\rho}\Gamma^\alpha_{\sigma\alpha}U_{,\rho}
+ \sqrt{-g}g^{\sigma\rho}\Gamma^\alpha_{\sigma\rho}U_{,\alpha}]\eqno{(17)}
$$
and $\hat {\rm B} \equiv {\rm B} + \sqrt{-g}L_{\phi + \psi}$, 
the conserved pseudo energy momentum vector defined over a hypersurface 
$\Sigma$ is:
$$
P_\mu \equiv \int_\Sigma d\Sigma[\sqrt{-g}T^o_{w\mu} 
- \hat {\rm B} \delta^o_\mu
- {{\partial\hat{\rm B}}\over {\partial g^{\tau\beta}_{,o}}}g^{\tau\beta}_{,\mu}
-{{\partial \hat{\rm B}}\over {\partial\phi_{,o}}}\phi_{,\mu}
-\bar\psi_{,\mu}{{\partial \hat{\rm B}}\over {\partial\bar\psi_{,o}}}
-{{\partial \hat{\rm B}}\over {\partial\psi_{,o}}}\psi_{,\mu}]
\eqno{(18)}
$$
This is also expressible as a surface integral over a 2 dimensional
boundary of $\Sigma$:
$$
P_\mu = -\int ({\partial\hat {\rm B}\over \partial
g^{\mu\nu}_{,j}}g^{0\nu})d\Sigma_j
\eqno{(19)}
$$
Thus in a generalised Brans-Dicke theory, the generalised energy
momentum is determined by the metric tensor and its derivatives on a
2-dimensional boundary.
\end{subsection}

\begin{subsection}*{2(c). The dynamic equations:}

We consider classical solutions to eqns.(6-9) for a fixed number of
fermions. As demonstrated in \cite{wick} and \cite{ho} the
thermodynamic potentials for the fermion field are assumed to be
adequately described in terms of the chemical potential and the
temperature. Denoting the fermion density by $S_f$, eqn(7) reads:
$$
\Box\phi + {\partial V\over {\partial\phi}} 
- R{\partial U \over {\partial \phi}} + 
{\partial m\over {\partial \phi}}S_f  
= 0 \eqno{(20)}
$$
The trace of eqn(6) gives:
$$
U(\phi)R = -(3U''(\phi) +{1\over 2})\phi^{,\alpha}\phi_{,\alpha}
+ 2V(\phi) + {1\over 2}mS_f - 3U'(\phi)\Box\phi \eqno{(21)} 
$$
Substituting in eqn(20) gives: 
$$ \Box\phi + {U'(\phi)(3U''(\phi)
+{1\over 2})\over {1 + 3{U'^2\over U}}} \phi^{,\alpha}\phi_{,\alpha} +
{{V' + m'S - {U'\over U}(2V + {1\over 2}mS)}\over {1 + 3{U'^2\over
U}}} \eqno{(22)}
$$
\end{subsection}

\begin{subsection}*{2(d). An example of tuning of $\Lambda$:}
Consider a familiar NMC theory with $U(\phi) = \beta/8\pi -\xi\phi^2/2$ and
$V(\phi) \equiv \Lambda_0/8\pi$ a constant. We define natural units in which
$\beta$ takes on a unit value.  The field equations
simply read:
$$
\Box\phi + \xi R\phi =0 \eqno{(23)}
$$
and
$$
G_{\mu\nu} + \Lambda_O g_{\mu\nu} = -8\pi T_{\mu\nu} \eqno{(24)}
$$
with
$$
T_{\mu\nu} ={\phi}_{,\mu}{\phi}_{,\nu} -1/2
g_{\mu\nu}{\phi}_{,\rho}{\phi}^{,\rho} - \xi{\phi}^2(R_{\mu\nu} -1/2
g_{\mu\nu}R) + \xi g_{\mu\nu}\Box{\phi^2} - \xi{\phi^2}_{;\mu;\nu} 
\eqno{(25)}
$$
In terms of $\Phi \equiv \phi^2$, the scalar field eqn(23) reads
$$
\Box{\Phi} = -2{\xi}R{\Phi} + {1\over {2\Phi}}
{\partial}_{\mu}{\Phi}{\partial}^{\mu}{\Phi} \eqno{(26)}
$$
while the trace of eq(24) is simply:
$$
- R  + 4\Lambda_0 = - 4\pi [- {\Phi_{,\rho}\Phi^{,\rho} \over {2\Phi}} +
2{\Phi}R\xi + 6\xi\Box{\Phi}] \eqno{(27)}
$$
The wave equation for the scalar field thus reduces to:
$$
\Box{\Phi} = {{4\Lambda_0 - R}\over {4\pi(1 - 6\xi)}} \eqno{(28)}
$$
We shall follow the dynamics of $\phi$ in a FRW metric:
$$
ds^2 = dt^2 - a^2(t)[{dr^2\over {1 - kr^2}} + r^2(d\theta^2 + 
sin^2\theta d\varphi^2)] \eqno{(29)}
$$
For a homogeneous, time dependent scalar field, the above equations reduce to:
$$
{3\over {8\pi}}[({\dot{a}\over a})^2 + {k\over {a^2}}] 
- {\Lambda_0\over 8\pi} = {\dot{\Phi}^2
\over {8\Phi}} + 3\xi[({\dot{a}\over a})^2 + {k\over {a^2}}]\Phi +
3\xi({\dot{a}\over a}){\dot{\Phi}} \eqno{(30)}
$$
and
$$
\ddot\Phi + 3(\dot{a}/a)\dot\Phi = {{4\Lambda_0 - R}\over {4\pi(1 - 6\xi)}} 
\eqno{(31)}
$$
We look for solutions with the scale factor having a power law
evolution: $a(t) = ht^\alpha$, with $h,~\alpha$ arbitrary
constants. The equation for $\Phi$ integrates exactly for all $\alpha
\ne 1/3$ in terms of integration constants $K_o,~C$:
$$
\Phi = At^2/2 + K_0 t^{1 - 3\alpha} + C \eqno{(32)}
$$
where
$$
A = {\Lambda_0 \over {\pi(1 - 6\xi)(1 + 3\alpha)}} \eqno{(33)}
$$
(For $\alpha = 1/3$ one has  $K_0ln(t)$ instead of the second term. From what
just follows, this is not feasible as a late time solution).  On
substituting in eqn(20) it is found that it is possible to have $k \ne
0$ solutions for large times only if $\alpha \ge 1$. For $\alpha > 1$,
the solutions are independent of the curvature constant (i.e. hold for all $k$). 
The scale $\alpha$ directly relates to the parameter $\xi$ as: $\xi =
-(4\alpha -2)^{-1}$. Similarly, for $\alpha = 1$, $\xi = -(2 +
2k/h)^{-1}$. For all these cases, scalar condensates develop to have
their energy cancell the effect of $\Lambda_0$:
$$
T_{\mu\nu} \longrightarrow - {1\over {8\pi}}\Lambda_0g_{\mu\nu} +
O(t^{-2}) \eqno{(34)}
$$
The cancellation occurs over time scales large as compared to the time
scales determined by dimensional parameters in the action. If one
merely uses typical particle physics time scales, this implies that
the stress energy of the scalar field would compensate the
cosmological constant for the entire history of the universe except
the very early history.  It is the coupling of the scalar field to the
spacetime curvature that is primarily responsible for the above
instability of the scalar field which in turn leads to an efficient
damping of the cosmological constant.  Unfortunately, the effective
value of the gravitational constant goes as $G_{eff} = (\beta + 8\pi
|\xi|\phi^2)^{-1}$.  This becomes unacceptably small at large
times. This is an example of a ``no-go'' situation \cite{wein2}.  This
behaviour was noted in \cite{dolgov,ford} for a flat $k = 0$ model. We
conclude that the presence of spatial curvature does not alter the
results. It is also straightforward to demonstrate that the results
are independent of the presence of the Planck sacle determined by
$\beta^{-1} \equiv 8\pi G_0 \equiv M_{plank}^{-2}$.  Indeed in the
absence of $\beta$ in the NMC function $U(\phi)$, the FRW scale
evolves as above: $a(t) \propto t^\alpha$ with $\alpha$ constrained as
before in the presence of $\beta$, the stress tensor rapidly cancells
the cosmological constant while the effective gravitational constant
again approaches zero over time scales large in comparison to the
particle physics time scales.

The above behaviour of a power law behaviour of the FRW scale factor
and the cancellation of the vacuum energy has been generalized to
scalar tensor theories within a large class of scalar field effective
potentials \cite{wett}. There have also been attempts to use the above
results to constrain dimensional parameters in the theory (see eg
\cite{chiba}).  One equates the observed gravitational constant to that
calcuted above using conservative age estimates of the universe.
Early universe nucleosynthesis constraints and recent Viking radar
experiments severly constrain the NMC in a scalar tensor theory. The
particle physics parameters that follow turn out to be rather
unnatural. Instead we propose and explore the possibility that the
above analysis successfully accounts for a NMC diverging with time and
that an effective gravitational constant over most of the universe is
indded vanishing. However, this is not the constant which ought to be
equated with the Newtonian gravitational constant.  Suspecting the
problem to lie in the assumption of homogeneity, it has been suggested
\cite{dolgov,ford} to look for a stabilization of $G_{eff}$ in an
inhomogeneous model.
\end{subsection}

\begin{subsection}*{2(e). A strategy for effective gravity:}

Taking a cue from the above, we consider a theory described by
eqns(2 \& 3).  We shall be interested in dynamic NMC's that diverge, as
described above, outside bounded domains. It is convenient to recast
the equations in terms of the effective planck length parameter which
goes to zero as the NMC diverges. One such reparametrization of the
scalar field is $\phi \longrightarrow M^2/\phi$. The NMC becomes
$U(\phi) \approx \beta - \xi M^4/\phi^{-2}$ \cite{wett}.  With this
reparametrization, the results of the last section translate to the
vanishing of the effective gravitational and cosmological constants
together with the vanishing of $\phi$ over time scales large in
comparison to any particle physics time scales in the theory. In other
words we consider an equivalent problem that has the NMC divergent at
any point that we take as $\phi = 0$ without any loss of
generality. Thus $U(\phi = 0) \longrightarrow \infty$. The model has a flat,
Minkowski spacetime as a fixed point global solution with $\phi = 0$
for an arbitrary equation of state of matter. However, there are
non-trivial solutions to the model as well \cite{dlms}.  We are
particularly interested in solutions that have $\phi \longrightarrow
0$ outside compact domains but not in the interior. This would mean
that the gravitational constant would vanish in the exterior. However,
this does not imply doing away with gravitation between such
domains. To see this, consider the Einstein tensor that follows from
equations (2) and (6):
$$
G_{\mu\nu} = [U(\phi)]^{-1}~~\times~~ source terms \eqno{(35)}
$$
From the expansion of the metric about a flat space metric:
$g_{\mu\nu} = \eta_{\mu\nu} + h_{\mu\nu}$, in appropriate 
(harmonic) coordinates, one gets: 
$$
\Box h_{\mu\nu} = [U(\phi)]^{-1}~~\times~~ source terms \eqno{(36)}
$$
The retarded solution reads:
$$
h_{\mu\nu}(x,t) = \int {d^3x'\over {|x-x'|}}
[U(\phi)]^{-1}~~\times~~ source terms(x', t - |x-x'|) \eqno{(37)}
$$
Thus as long as the effective gravitational constant at the
retarded source  point $(x', t - |x-x'|)$ is bounded, the vanishing
of the effective gravitational constant at the field point is 
of no consequence for the above solution. All one has to 
discover is the means of having the NMC locked to a universal
value in non-trivial domains. The  mechanism we invoke is
to have the non-minimally coupled scalar field locked at the minimum 
of its effective potential \cite{Zee}.
\end{subsection}

\begin{subsection}*{2(f). Large Non trivial  solutions:}

We recall studies on non-trivial configuration that have $\phi$
vanishing outside, and $N_f >> 0$ $\psi$ fermions trapped inside,
compact domains in the Lee-Wick model \cite{wick}. With the fermion
number conserving, the minimum energy configuration of such solutions
are non-topological soliton solutions [NTS's] in the theory. We recite
the essential properties of such an NTS. Lee and Wick considered a
theory with a Lagrangian:
$$
L = {1\over 2}\partial_\mu\phi\partial^\mu\phi - V(\phi)
+ \bar\psi\gamma^\mu\partial_\mu\psi 
- m\bar\psi[1 - {\phi\over \phi_0}]\psi \eqno{(38)}
$$
where
$$
V(\phi) = {1\over 2}m_\phi^2\phi^2[1 - {\phi\over \phi_0}]^2
\eqno{(39)}
$$
With Fermion number conserved in this theory, solutions representing a
degenerate distribution of fermions with a total fermion number $N_f$,
at $\phi = \phi_0$, would remain confined there if the total energy of
a fermion is less than its on-shell energy at $\phi = 0$. The NTS is a
ball of radius $r_b$ inside which the scalar field is locked at $\phi
\approx \phi_0$.  Across the surface of the ball of thickness $\sim
m_\phi^{-1}$, the scalar field transits to $\phi = 0$. As a matter of
fact, the spherical shape of the distribution follows from degeneracy
of the fermion ensemble. For a degenerate fermion distribution, the
fermion density is determined in terms of the chemical potential that
occurs as a lagrange multiplier for the conserved fermi number of the
ensemble. This determines the volume for a fixed conserved number of
degenerate fermions. A positive definite surface energy would break
the degeneracy of volume preserving (i.e. fermion number conserving)
deformations of the distribution
- choosing a configuration of minimum surface area. This gives the
minimum energy configuration to be a sphere.  In terms of the chemical
potential $\mu$, the fermion density goes as $\rho_f \sim \mu^4$. The
fermion number density goes as $n_f \sim \mu^3$. With the total energy
and fermion number: $E_f = \rho_fv \sim \mu^4r_b^3$, $N_f = n_fv \sim
\mu^3r_b^3$, the chemical potential eliminates to give $E_f \sim
N_f^{4/3}/r_b$.  For a large enough $r_b$, the surface tension is
given by the expression: $s \sim m_\phi\phi_0^2/6$, the total surface
energy being $E_s = 4\pi sr_b^2$. For a given $N_f$, one may vary the
radius to seek the minimum of the volume and surface energies to give
$E_f = 2E_s$. The mass of the soliton is expressed as $M = E_f + E_s =
3E_s = 3E_f/2 = 12\pi sr_b^2$. This allows one to express:
$$
N_f \sim s^{3/4}r_b^{9/4}; ~~~M \sim s^{1/3}N_f^{8/9};~~~
r_b \sim s^{-1/3}N_f^{4/9} \eqno{(40)}
$$
Thus with a given surface tension, one could get arbitrarily massive
NTS's by increasing the fermion number. However, the above flat
spacetime analysis can not be extended past the Schwarzschild limit.
In the Lee-Wick model, the Schwarzschild limit can be determined as the
criteria for stability against gravitational collapse.  As a matter
of fact, gravitational instability occurs even before this limit is
reached \cite{co}. We shall use the Schwarzschild limit as a
benchmark to keep sufficiently clear-off and as a convenience that ensures 
that the flat spacetime analysis holds.  This limit is simply obtained
by equating the radius of the NTS to the Schwarzschild radius for a
critical mass ball $M_c$: equating $r_b = 2GM_c$, and with the
expression for the mass of a ball, one gets $M_c \sim (48\pi
G^2s)^{-1}$. For $s \sim (30 Gev)^3$, $M_c$ is roughly
$10^{15}M_\odot$ with the critical radius some $10^2$ Pc.

One can have NTS's as large as some 10's of kilo parsecs.  This would
require a class of low mass fermions that could keep the NTS from
collapsing.  In a hot big bang cosmology, any species of particles
that decoupled from equilibrium when it was relativistic, ought to
have the same relic density as say the relic photons or neutrinos:
some $\sim 200$ per cc. To get $\sim 10$ to $100$ KPc NTS's, it would
suffice to have $s \sim (Mev)^3$. This gives $N_f \sim 10^{72}$ to
$10^{75}$ with the mass of the ball some 12 orders of magnitude
smaller than the the Schwarzschild (critical) mass for $s \sim
(Mev)^3$. We conclude that for such a surface tension and for NTS's
even as large as hundreds of kilo parsecs, one can consistently ignore
the spacetime curvature in such solutions. In the following, we do
this whenever convenience demands.

In the model described by eq(3), fermions $\psi$ are confined to
$\phi \ne 0$ domains. It is easy to follow \cite{meetu,dlms} to
demonstrate the existence of non-trivial solutions.  In the ``weak
(gravitational) field approximation'', that would justify retaining
only a first order deviation from a flat metric, the metric can be
expressed in terms of the spherical [Schwarzschild] coordinates:
$$
ds^2 = e^{2u}dt^2 - e^{2\bar v}dr^2 - r^2[d\theta^2 + 
sin^2\theta d\varphi^2] \eqno{(41)}
$$
We look for a static solution describing the scalar field trapped to a
value $\phi = \phi_{in}$ in the interior of a sphere of radius $r_b$
and making a transition across a thin surface to $\phi = 0$
outside. The fermi gas trapped inside the soliton is described, in the
Thomas Fermi approximation \cite{wick}, by the following
distribution in momentum space: $n_k = \theta (k - k_f)$, $k$ being
the momentum measured in an appropriate local frame that depends on
$r$, and $k_f$ the fermi momentum.  The fermion energy density is
given by:
$$
W = {2\over 8\pi^3}\int d^3kn_k\epsilon_k\eqno{(42)}
$$
with $\epsilon_k = \sqrt{k^2 + m(\phi_{in})^2}$. The fermion number
density $\nu_f$ and the non-vanishing components of fermion stress 
energy tensor are:
$$
\nu_f = {2\over 8\pi^3}\int d^3kn_k\eqno{(43)}
$$
$$
T^t_t = W
$$
$$
T^r_r = T^\theta_\theta = T^\varphi_\varphi = T^\rho_\rho
\equiv -T = -{2\over 8\pi^3}\int d^3kn_k{k^2\over 3\epsilon_k}\eqno{(44)}
$$
The trace of the stress tensor is just:
$$
T^\mu_\mu = W - 3T = m(\phi_{in})S\eqno{(45)}
$$
with S the scalar density:
$$
S = {2\over 8\pi^3}\int d^3k{n_k\over \epsilon_k}m(\phi_{in})\eqno{(46)}
$$
Defining $8\pi G_{in} \equiv U(\phi_{in})^{-1}$ as the effective
interior ``gravitational constant'', the metric field equation in the 
interior can be expressed in the above spherical coordinates as:
$$
r^2G^t_t = e^{-2\bar v} - 1 - 2 e^{-2\bar v}r{d\bar v\over dr}
= - 8\pi G_{in}r^2[W + V(in)]\eqno{(47)}
$$
$$
r^2G^r_r = e^{-2\bar v} - 1 + 2 e^{-2\bar v}r{du\over dr}
= 8\pi G_{in}r^2[T - V(in)]\eqno{(48)}
$$
$$
r^2G^\theta_\theta = e^{-2\bar v}[r^2{d^2u\over dr^2} 
+ [1 + r{du\over dr}]r{d\over dr}(u - \bar v)]
=  8\pi G_{in}r^2[T - V(in)]\eqno{(49)}
$$
The scalar field satisfies:
$$
\phi^{;\mu}_{;\mu} + V' + m'(\phi)S - U'R = 0 \eqno{(50)}
$$
Taking the trace of the Einstein  tensor $G^\mu_\mu$, the
Ricci scalar $R$  substitutes in eqn(24) to give the following
radial equation for the scalar field in the weak
field limit:
$$
{d^2\phi\over dr^2} + {2\over r}{d\phi\over dr} 
+ F(U)[{d\phi\over dr}]^2 = {dW\over d\phi} \eqno{(51)}
$$
where
$$
{dW\over d\phi} \equiv [V' + m'(\phi)S - {U'\over U}({mS\over 2} + 2V)]/
[1 + {3U'^2\over U}] \eqno{(52)}
$$
and 
$$
F(U) \equiv  {U'\over U}({1\over 2} + 3U'')/
[1 + {3U'^2\over U}] \eqno{(53)}
$$
A sufficient condition for existence for a large NTS with $\phi_{in}
\sim$ constant is the vanishing of $W'$ at $\phi_{in}$. A convenient 
way to arrange this is to have $V(\phi)$ to have a minima at $\phi_{in}$,
where $U'/U$ is small, and to have fermions massless at $\phi_{in}$.  
The zero of $W'$ then coincides with that for $V'$. For a profile of 
$V$ described in Fig(1), and $U(\phi) = M_p^2 + M^4/\phi^2$ 
(with $\phi^2 >> M^4/M_p^2 $), the profile of $W$ closely follows that
of $V$ except near $\phi \approx 0$. The divergence of the NMC at $\phi =0$
ensures the vanishing of $W'(\phi = 0)$. The divergence of the NMC and in turn
that of $F(U)$ also ensures that $\phi = 0$ is a solution with the Ricci
scalar vanishing for arbitrary $V(\phi = 0)$. A typical non-trivial NTS 
solution has the scalar field locked to $\phi = 0$ outside a compact spherical
domain of radius $r_b$ and transits to $\phi = \phi_{in}$ across the boundary:
$$
1 - {\phi\over \phi_{in}} \longrightarrow exp[-V''(\phi_{in})(r_b - r)]
$$
The transition zone is thus related only to $V''(\phi_{in})$ for large 
$r_b$. The radius $r_b$ is determined by the energetics of the ball and
is governed by the number of trapped fermions inside the ball.
The pressure of the degenerate fermions keeps the soliton from
collapsing. As for the Lee-Wick NTS (see also \cite{co}) the total energy 
is expressible as the sum of (i) the volume energy, (ii) the surface
energy and (iii) the energy of the fermions:
$$
E = {4\pi\over 3}V(\phi_{in})r_b^3 + 4\pi sr_b^2 
+ \alpha  {N_f^{4/3}\over r_b}; ~~~
\alpha \equiv {1\over 2}({3\over 2})^{5/3}\pi^{1/3} \sim O(1)
\eqno{(54)}
$$
For a vanishing $V(\phi_{in})$, the minimum of (ii) and (iii) give the
total energy $E \sim N_f^{8/9}(4\pi s)^{1/3}$, with an energy per
fermion $\sim (4\pi s)^{1/3}/N_f^{1/9}$. For $V(\phi_{in}) > 0$, large
$r_b,~N_f$ solutions have energy $E \sim 4(4\pi V)^{1/4}N_f/3$, with
the energy per fermion roughly independent of the number of
fermions. Finally, for negative $V(\phi)$, NTS's exist only for a
small fermion number $(\bar N_f)$ and for small enough $|V|$. In this last case,
the existence of NTS's requires the surface term to dominate over the
volume term. This is possible only for small $r_b,~\bar N_f$. The
energy per fermion is again $\sim (4\pi s')^{1/3}/\bar N_f^{1/9}$.
{\it Energy per fermion for a large fermion number configuration is
smaller, the closer the minimum of $|V|$ is to zero.}

This can form the basis for a prescription to sort out the
cosmological constant problem. We first recall that in standard GTR,
the cosmological consatnt problem is sought to be {\it solved} by
first making the cosmological constant dynamical: i.e. a constant of
motion.  This is done by using (for example) Witten's \cite{witten}
prescription by Coleman, Hawking and Weinberg
\cite{wein2,hawking,cole}.  One adds to the action a term:
$$
I_F = -{1\over 48}\int d^4x\sqrt{-g} F_{\mu\nu\rho\sigma}F^{\mu\nu\rho\sigma}
$$
with $F_{\mu\nu\rho\sigma}$ the exterior derivative of a three form gauge
field $A_{\nu\rho\sigma}$:
$$
F_{\mu\nu\rho\sigma} = \partial_{[{\mu}A_{\nu\rho\sigma}]}
$$
using the fact that $F_{\mu\nu\rho\sigma}$ is totally antisymmetric, it
is proportional to the permutation tensor.  The field  
equation for $A_{\nu\rho\sigma}$ implies that the proportionality factor is a 
constant. This reduces the action to:
$$
I_F = {1\over 2}c^2\int d^4x \sqrt{-g}
$$
With $c$ a constant of integration. This makes the effective
cosmological constant a constant of integration. This being the case,
then in a quantum theory we expect the wave function of the universe
to be a superposition of states with different values of the
cosmological constant. Hawking \cite{hawking} has proposed that the
application of conditional probability that homosapiens would come
into being and study physics in a universe then {\it solves} the
cosmological constant.  However, such an anthropic principle is not a
dynamical principle in the first place.  Even if the evolution of an
intelligent species were related to a dynamic prescription, it would
be difficult to implement it in standard GTR where every value of an
effective cosmological constant corresponds to a distinct FRW solution
that can not be dynamically compared with a solution with a different
cosmological constant.

In the structure that we have described, the situation is simpler. One
can start with the action eqn(2) and consider NTS's with a large
average fermion density in the universe. Every NTS is a ball that has
an effective gravitational constant vanishing in its exterior and
having a unique value in its interior determined by the NMC at
$\phi_{in}$ where the effective potential is minimized. The effective
cosmological constant vanishes in the exterior and the value of
$V(\phi_{in})$ contributes to the interior volume energy of the
NTS. One can now play the same game as played in standard GTR and
introduce $I_F$ as before to make $V(\phi_{in})$ dynamical. Consider
the wave function of the universe containing a large (conserved)
number of fermions. We can expect the state vector of the universe to
be a superposition of states with different $V(\phi_{in}$). From the
previous considerations, the minimum energy per fermion would be for
the state that has $V(\phi_{in})$ vanishing. The minimization of
energy is a dynamic principle that may or may not have anything to do
with the existence of apes !! With all NTS's having having common
asymptotics for an arbitrary $V(\phi_{in})$, a dynamical solution that
uses an energy minimization prescription makes sense and is easy to
implement.
\end{subsection}

\begin{subsection}*{2(g) Microscopic domains:}

It is interesting to note that were we to ensure bounded NMC inside
domains as large as hadrons, one would again get an effective gravity
model with no cosmological constant problem. Consider for example a
theory described by eqn(42) in which the confined fermionic degrees of
freedom were associated with baryons. With the NMC diverging at say
$\phi = 0$ and the fermions confined to hadronic size microscopic $\phi \ne 0$
domains, there would be effective gravitational effects for baryons. A
``cosmological constant'' as large as $\Lambda_{QCD} \approx 
~(.2 - .3Gev)^4$ would merely make a small contribution to the mass of a
baryon and could be absorbed in its net mass. The vanishing of the
covariant divergence of the matter stress energy tensor implies the geodesic 
law of motion for material particles. The geodesic law in turn implies
the equality of the inertial mass and the passive gravitational
mass. The gravitating (active) gravitational mass, however, would
depend upon the number and binding energy of non-baryonic matter of
the source. Considering the relatively large error bars in the
determination of the gravitational constant in a typical Cavendish
experiment, the change in the active gravitational mass would not be
discernible. However, it is claimed \cite{will} that the equality of
the active and passive gravitational masses is established in lunar
ranging experiments to a very high accuracy.  For this reason (not
withstanding our doubts on the claimed accuracy and interpretation of
the above experiments) we feel that the large NTS's described earlier
offer a better promise for a problem free gravity theory.
\end{subsection}

\end{section}

\begin{section}*{3. Features of a toy cosmology:}

In the model described above, a divergent NMC over most of the
universe provides for a vanishing cosmological constant. Non-trivial
solutions for the scalar field provides for gravitating pockets inside
compact domains - with the divergent NMC being restricted to their
exterior.  A value $N_f \approx 10^{75}$ for $s \approx (Mev)^3$ would
give a NTS of a size of tens of kilo parsecs. Such an $N_f$ is of the
same order as the relic background neutrinos / photons in the
universe. Thus a fermion species that decouples very early in the
universe when that species is still relativistic, 
would be sufficient to provide $N_f$ for gravitating domains
as large as a Halo of a typical large structure (galaxy / local group
etc.).

With the effective gravitational constant identically vanishing
outside a NTS, one could conceive of a cosmological model that starts
with a hot big - bang and evolves as a Milne model \cite{milne} almost
from its birth. We assume the existence of conditions conducive for
the production of above NTS's with fermions confined to compact
domains where the scalar field is non-vanishing. These NTS's would
then evolve by mutually colliding and coalescing. As the NTS's
coalesce, the fermion number adds up, the NTS becoming larger and
could evolve to the current distribution of gravitating domains.

On large scales, the universe evolves with the Freidmann - Robertson -
Walker scale factor increasing linearly with time: $a(t) \propto
t$. This has characteristic features: (i) With $\int_0^t dt/a(t)$
unbounded for any $t > 0$, there is no horizon problem in the
theory. (ii) With the expansion parameter not determining any
``critical - density'' in the model, there is no flatness (fine
tuning) problem.  (ii) The model is concordant with standard classical
cosmological tests, viz.: the number count, angular diameter and the
luminosity distance variation with redshift.  The first two tests are
quite sensitive to models of galactic evolution and for this reason
have (of late) fallen into disfavour as reliable indicators of a
viable model. However the magnitude - redshift measurements on SN 1A
have a great degree of concordance with $\Omega_\Lambda = \Omega_M =
0$ (\cite{dlms,kaplin,abha,perl}). (iii) With the scale factor
evolving linearly with time, the Hubble parameter is precisely the
inverse of the age t. Thus the age of the universe inferred from a
measurement of the Hubble parameter is 1.5 times the age inferred by
the same measurement in standard matter dominated model. Such a
cosmology promises consistency with an older universe.  (iv) The
deceleration parameter is predicted to vanish. (v) early universe
nucleosynthesis is not ruled out in a Milne universe. It has been
shown \cite{nuc} that for a baryon entropy
ratio $\eta \approx 5\times 10^{-9}$, one gets $\sim 24\%$ of helium-4
and metallicity quite close to that observed in type II stars and low
metallicity clouds. The cosmology thus comes with its characteristic
predictions and may well be distinguishable by the next generation of
experiments.
\end{section}

\begin{section}*{4. Conclusion:}

What we have profiled is a plausible program with an exploratory
spirit. We had been looking for a framework that could provide for a
spatial variation of the effective gravitational constant as a
solution to the cosmological constant problem. It was this search that
attracted our attention to remarkably encouraging results in the
theory of non - topological solitons [NTS's] in the Lee-Wick
model. Given a (large), conserved fermion number and a given surface
tension of such a soliton, the smaller the magnitude of the interior
volume energy density (given by the value of the minimum of the
effective potential), the smaller is the energy per fermion of the
NTS. If one adds non-minimal coupling, that diverges at a point that
we take (without any loss of generality) as $\phi = 0$, and makes the
interior volume energy dynamical, the minimizimg energy can provide a
dynamic prescription to sort out the cosmological constant
problem. The effective gravitational constant also approaches a
universal value determined by the NMC at the minimum of the effective
potential.

We have demonstrated that in a whole class of scalar tensor theories
in which the non-minimal coupling diverges and for which the classical
effective potential vanishes at some point, classical scalar field
condensates can occur as NTS's.  The effective gravitational constant
inside all large domains would approach a universal value and the
effective cosmological constant would drift to zero. The dynamical
tuning of the effective cosmological constant to a small value and the
effective gravitational constant to a universal value are compelling
features - enough to explore the possibility of raising the toy model
described here to the status of a viable cosmology.

\vskip 1cm
\centerline{Acknowledgements:}

This work has benefitted greatly from discussions with
Profs. T. W. Kibble, M. Rocek, P.D. Mannheim, L. Parker and
G. W. Gibbons.
\end{section}

\vspace{2cm}
\begin{center}{\bf Figure caption:}
\end{center}
\vspace{1cm} 
Figure 1:The profile of the potential, V and the effective potential,
W. The two can be chosen to 
match for all $\phi > 0$ except for $\phi$ near zero.
\end{document}